\documentclass[preprint2,usenatbib]{mnras}
\usepackage{graphicx}
\usepackage {amsmath,amssymb}
\usepackage{color}
\usepackage{aas_macros}
\usepackage{array}
\usepackage{booktabs}

\newcommand{\Msun}{{\rm M}_{\odot}}
\newcommand{\Mstar}{{M}_{*}}
\newcommand{\krot}{\kappa_{rot}}
\newcommand{\jstar}{{j}_{*}}

\usepackage[dvipsnames]{xcolor}

\title[The emergence of the faint nature of Low Surface Brightness Galaxies]{The emergence of the faint nature of Low Surface Brightness Galaxies in the IllustrisTNG simulation}

\author[L. E. P\'erez-Monta\~no, et al.]
{
	\parbox{18cm}{
		Luis Enrique P\'erez-Monta\~no,$^{1}$\thanks{E-mail: le.perezmontano@zju.edu.cn}
        Bernardo Cervantes Sodi,$^{2}$\thanks{E-mail: b.cervantes@irya.unam.mx}
        Vicente Rodriguez-Gomez,$^{2}$ Doris Stoppacher,$^{3}$ and Tian-Wen Cao$^{4,5}$
	}
	\vspace{0.3cm} \\ 
    $^{1}$ Institute for Astronomy, School of Physics, Zhejiang University, Hangzhou 310027, People’s Republic of China\\
	$^{2}$ Instituto de Radioastronomía y Astrofísica, Universidad Nacional Autónoma de México, C.P. 58089, Morelia, Michoacán, México \\
    $^{3}$  Facultad de Físicas, Universidad de Sevilla, Avda. Reina Mercedes s/n, Campus de Reina Mercedes, 41012 Sevilla, Spain \\ 
    $^{4}$ National Astronomical Observatories, Chinese Academy of Sciences, Beijing 100101, People’s Republic of China\\
    $^{5}$ Guizhou Radio Astronomical Observatory, Guizhou University, Guiyang 550000, People’s Republic of China \\
}

\date{Accepted XXX. Received YYY; in original form ZZZ}

\pubyear{2025}

\begin{document}
\label{firstpage}
\pagerange{\pageref{firstpage}--\pageref{lastpage}}
\maketitle

\begin{abstract}
We employ a simulated sample of galaxies drawn from the IllustrisTNG suite to study the emergence of the diffuse and extended nature of $\sim12,000$ low surface brightness galaxies (LSBGs) within a wide stellar mass range ($\Mstar=10^{9}-10^{12} \Msun$). We employ merger trees to follow the evolution of their physical properties such as stellar surface density, specific angular momentum and halo spin parameter, finding that the central low density nature of LSBGs is mainly a consequence of an increase in their angular momentum and (inner) halo spin parameter. We also find that star formation histories of LSBGs are quite similar to their high surface brightness (HSBGs) counterparts, with significant differences not in the time, but in the spatial distribution in which new stars are forming. We conclude that the mechanisms that favor the emergence of the low surface brightness nature are strongly related with variations in the spin parameter of host halos and their angular momentum, deviating the stellar distribution of galaxies from their inner regions to their outskirts, leading to a decrease in their central surface brightness. Once the LSBG nature is established, galaxies are less likely to experience strong variations in their central surface densities and morphology.

\end{abstract}

\begin{keywords}
galaxies: fundamental parameters -- galaxies: formation -- galaxies: haloes -- galaxies: statistics 
\end{keywords}

\section{Introduction}
\label{sec:Intro}

An early study of disk galaxies by \citep{Freeman70} found that most galaxies had a similar central surface brightness of $\mu_B \sim$ 21.65 mag arcsec$^{-2}$, regardless of their absolute magnitude. However, this study also suggested the presence of galaxies with central surface brightness below this limit, a finding later highlighted by \citet{Disney76}. Galaxies with central surface brightness lower than this limit are commonly referred to as `Low Surface Brightnes' galaxies (LSBGs). Later observations \citep{McGaugh95a} confirmed the existence of such galaxies and demonstrated that they represent a significant fraction of extragalactic sources, making up about $\sim$ 30-50\% of the galaxy population \citep{McGaugh95a,ONeil03}. Advances in observational techniques, new extragalactic surveys, and more sensitive instruments have further increased this fraction, suggesting that LSBGs could represent a substantial part of the Universe's baryonic budget.

Over the years, many authors have proposed various definitions to classify LSBGs, depending on different factors such as the aperture used to measure the flux of their central regions \citep{Boissier03,Galaz11,Kulier20,Stoppacher25}, the initial assumptions about their light profiles \citep{Sarkar25}, and the photometric band used to characterize their stellar population \citep{Zhong08, Bakos12, Tanoglidis21}. Nevertheless, the classical $\mu_B \sim$ 21.65 mag arcsec$^{-2}$ threshold remains widely used in the literature to distinguish LSBGs from their `High Surface Brightness' (HSBGs) counterparts. 

A key difference between LSBGs and HSBGs is that LSBGs reside within dark matter halos with high angular momentum, reflected in their high spin parameter ($\lambda$). This is attributed to tidal forces from neighboring overdensities during the early stages of the Universe \citep{Hoyle51,Peebles69}. \citet{Jimenez98} and \citet{McGaugh21} suggest that galaxy properties such as surface brightness are influenced by their spin parameters. A variation in $\lambda$ between 0.04 and 0.1 can cause a 2 mag change in surface brightness above the threshold set by \citet{Freeman70}, a result confirmed by both theoretical \citep{KimLee13,DiCintio19,Zhu23} and observational studies \citep{Dalcanton97,Boissier03,LEPM19,Salinas21}. Thus, the high angular momentum and spin parameters of LSBGs are considered the primary factors behind their low stellar densities and surface brightness.

An intriguing question regarding to their nature is whether they originated as LSBGs or underwent transformations over cosmic time, affecting their surface brightness profiles, sizes, and the presence of rotationally-supported structures \citep{Abraham01,DiCintio19,Park22}. If such transitions occur, it remains unclear whether they result from gradual internal changes, such as variations in angular momentum and spin parameter, external factors like tidal interactions and mergers, or a combination of both. In addition to star formation histories, surface brightness and morphology, other galaxy properties such as accretion history and the growth of supermassive black holes, which has been proven to be linked to changes in angular momentum and spin parameter \citep{RodGom22,Singh23}, also evolve. 

Recent studies, such as \citet{DiCintio19}, using a sample of 12 galaxies from the NIHAO zoom-in simulations \citep{Wang15}, suggest that mergers play a significant role in shaping the surface brightness of galaxies at $z=0$. This effect is not related to merger ratios or timings, but rather to orbital parameters and merger configurations. Coplanar co-rotating mergers, in particular, tend to increase galaxy angular momentum, leading to a decrease in surface brightness. This result was later confirmed by \citet{Wright25} using simulated galaxies from the ROMULUS25 simulations \citep{Tremmel17}. Similarly, \citet{Wu25}, using the TNG100 run of the IllustrisTNG project \citep{Nelson18, Pillepich18, Springel18}, pointed out the impact of orbital configurations on galaxy morphology. On the other hand, \citet{Martin19} found no significant differences in the angular momentum distributions of LSBGs and HSBGs progenitors in the HorizonAGN simulation \citep{Dubois14}, suggesting that the formation of LSBGs is not primarily due to their position in the high-spin tail of the angular momentum distribution, but rather external processes like tidal perturbations and ram-pressure stripping, which dominate their evolution at $z<1$.

By contrast, other studies emphasize that secular processes, rather than environmental effects, primarily drive the evolution of LSBGs, especially those related to angular momentum and spin parameters. \citet{LEPM22}, using a sample of galaxies from the IllustrisTNG simulations, found that LSBGs and HSBGs progenitors follow similar evolutionary paths at early epochs, but by $z\sim2$, the median value of $\lambda$ for LSBGs diverges towards higher values compared to HSBGs. Additionally, the evolutionary tracks of angular momentum and galaxy size diverge between LSBGs and HSBGs at lower redshifts, suggesting a cause-and-effect relationship between spin parameter evolution and other galaxy properties. This trend was later confirmed by \citet{Ma24} and \citet{Tang24}. Similarly, using a sample from the EAGLE simulations \citep{Schaye15, Crain15, McAlpine16}, \citet{Stoppacher25} concluded that the low surface brightness of galaxies is mainly due to intrinsic dynamical and structural parameters, with angular momentum being the key factor driving the divergence in evolutionary paths at high redshifts.

An important factor that seems to be crucial for the `survival’ of LSBGs is the isolated nature of their surrounding environment. Both observational \citep{Bothun93, Rosenbaum09, Galaz11} and theoretical \citep{Saburova23, Zhu23} studies show that LSBGs are generally farther from their closest neighbors than HSBGs, with fewer galaxies in their vicinity. \citet{Kulier20}, using a sample from the EAGLE simulations, suggested that LSBGs are not formed in low-density environments, but rather become isolated due to the accretion of nearby galaxies that form diffuse, extended disc-like structures around them. In \citet{LEPM24}, it was shown that the two-point correlation function of LSBGs in TNG100 has a lower amplitude than HSBGs, particularly at scales below 1 Mpc, indicating that LSBGs are more isolated. These studies highlight that the local environment (within 1 Mpc) and the galaxy's own halo play a crucial role in determining the observed properties of LSBGs.

This paper is the third part of a detailed study on the LSBG population in TNG100, following \citet{LEPM22} and \citet{LEPM24}. It aims to explore the changes experienced by LSBGs and HSBGs progenitors over cosmic time, identifying key parameters that either gradually or suddenly increase galaxy surface brightness, as well as other physical properties that influence the formation of LSBGs whether through a single event or multiple transitions. We also investigate whether these changes are driven by secular or environmental phenomena. The paper is organized as follows: In Section \ref{sec:Sample}, we describe the simulated sample employed in this work. In Section \ref{sec:Results}, we present the main results of our work, describing the evolution of different physical parameters of galaxies in our sample, in Sections \ref{sec:Discussion} we present a detailed discussion of the obtained results and how they compare with the existing literature within the formation and evolution of LSBGs framework, and finally Section \ref{sec:Conclusions} presents a brief summary of the current manuscript and its main conclusion.

\section{Simulated Sample}
\label{sec:Sample}

    \begin{figure*}
        \centering
        \includegraphics[width=\textwidth]{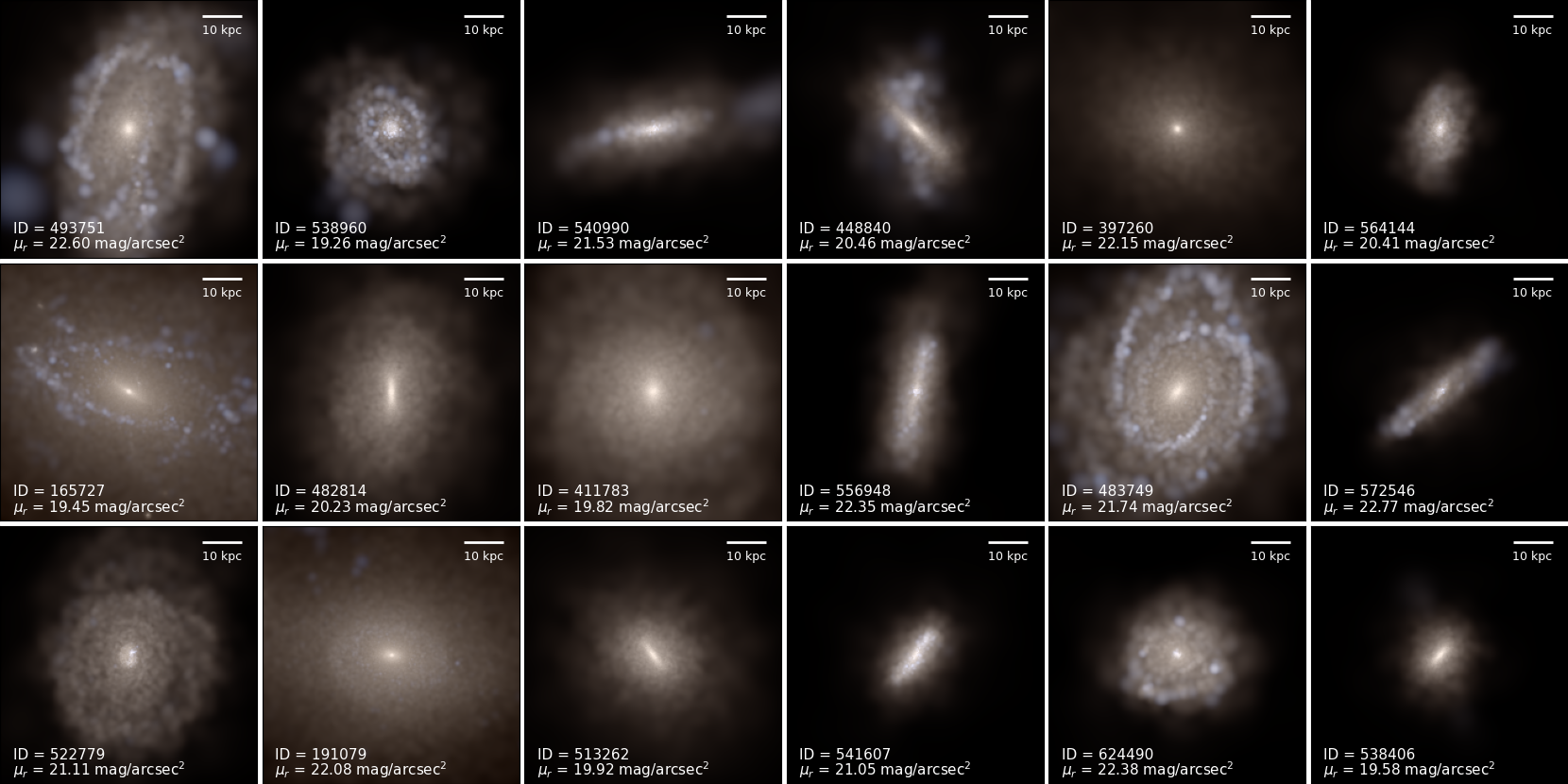} \\
        \caption{Idealized synthetic images of eighteen randomly selected galaxies in our sample with different surface brightness and sizes, highlighting that LSBGs and HSBGs in TNG100 exhibit a large variety of sizes and morphologies, as previously found in \citet{LEPM22}. These images were obtained following the \texttt{galaxev} pipeline by \citet{RodGom19} applied to the $g,r,i$ bands of the Subaru Hyper Supreme-Cam (HSC), projected on the $xy$ plane.}
        \label{fig:images}
    \end{figure*}

\subsection{The IllustrisTNG simulation}

The Illustris project \citep{Vogelsberger13,Genel14,Torrey14,Vogelsberger14,Nelson15} is a set of state-of-the-art cosmological, hydrodynamical simulations that follows the dynamics of gas coupled with dark matter with a quasi-Lagrangian treatment employing the moving-mesh code \texttt{AREPO} \citep{Arepo10}, allowing an adaptative dynamical discretization in which the points of the mesh that solve the hydrodynamical equations are moving together with the gas. The simulation employs a Friends-of-Friends algorithm (FoF, \citealt{Davis85}) to identify the dark matter halos with a minimum of 32 particles and a linking length $b=0.2$. Each halo contains subhalos, identified with the \texttt{SUBFIND} algorithm \citep{Dolag09,Springel01}, and each subhalo has a well-defined mass based on the particles gravitationally bounded to it. 

The upgraded version of the `original' Illustris galaxy formation model, named IllustrisTNG (hereafter TNG, \citealt{Nelson18,Pillepich18,Springel18}) is a set of 18 different cosmological simulations, 9 N-body and 9 magneto-hydrodynamical, which vary in size and resolution level, including new physical processes such as magneto-hydrodynamics, new prescriptions for black hole formation, growth and multimode feedback, stellar evolution and chemical enrichment \citep{Pillepich18b}. Three different box sizes of approximately 50, 100 and 300 Mpc per side are employed, labeled as TNG50, TNG100 and TNG300 respectively, each one with 100 snapshots running from $z=20$ to $z=0$, allowing to study galaxy formation at different scales. Each box runs three different simulations at different resolution levels: 2$\times$ 455$^3$, 910$^3$ and 1820$^3$ particles. The TNG model is based on a cosmology drawn from the Planck Collaboration \citet{Planck16} in which $\Omega_m=$ 0.31, $\Omega_\Lambda=$ 0.69, $\Omega_b=$  0.0486 and $h=$ 0.677. Further analysis in this paper will keep this convention. 

\subsection{Sample Construction and Selection Criteria}
\label{sec:sample_const}

We employed the TNG100 run at its highest resolution level (hereafter, TNG100-1) from the snapshot corresponding to $z=0$. We consider only those galaxies with stellar masses above 10$^9 \Msun$, in order to avoid spurious objects with very few stellar particles that define a disk, allowing a good matching with observational samples (e.g., \citealt{Galaz11,LEPM19}). Segregation of galaxies between nowadays LSBGs and HSBGs follow the same procedure as in \citet{LEPM22} and \citet{LEPM24}, which we briefly describe in the following lines.
    
For each subhalo, we calculate the specific angular momentum of the stellar particles $\vec{\jstar}$ and employ the azimuthal component to project each particle position vector along the $z-$axis (corresponding to a `face-on' projection). Then we calculate the central surface brightness following \citet{Zhong08} and \citet{Bakos12}, that is

    \begin{equation}
	\label{eq:mu_x}
	   \mu = m +2.5 \log{(\pi R_{50}^2)},
    \end{equation}
    
\noindent where $m$ is the apparent magnitude of the galaxy measured considering all the stellar particles within the effective radius $R_{50}$, which contains half of the total luminosity on a specific band, and has units of arcsec. Apparent magnitudes are calculated assuming that galaxies are located at $z=0.0485$ (corresponding to snapshot 95), which allows to match synthetic images \citep{RodGom19} with galaxies surveys such as SDSS \citep{York00} and Pan-STARRS \citep{Chambers16}. Luminosities in TNG are obtained following a \citet{BC03} single stellar population model, whereas $R_{50}$ is drawn from a linear interpolation over the cumulative distribution of particle luminosities as a function of their projected distance $R$ to the center of the corresponding subhalo. 
   
Traditionally, the segregation between LSBGs and HSBGs is done according to the selection criteria from \citet{Freeman70}, however it is always possible to perform this segregation employing other filters centered at different wavelengths (e.g., \citealt{Ragaigne03,Tanoglidis21,Ruiz25}). Recent observational \citep{Zhong08,Bakos12} and theoretical \citep{DiCintio19,Tang24} studies have adopted the SDSS $r-$band as their selection criteria, given that it allows to perform a good mapping of the ``underlying'' stellar population rather than Johnson's $B-$band, frequently related to their recent star formation activity. In this study, and following the procedure of previous studies in this series \citep{LEPM22,LEPM24}, we employ $\mu_r = 22.0$ mag arcsec$^{-2}$ as our selection criteria, so that galaxies with central surface brightness $\mu_r > 22.0$ mag arcsec$^{-2}$ are considered as LSBGs, otherwise, galaxies are classified as HSBGs. Fig. \ref{fig:images} shows a set of idealized images of eighteen randomly selected galaxies in our sample, obtained with the \texttt{galaxev} pipeline from \citet{RodGom19}. No dust attenuation is considered in this calculation, following \citet{Kulier20} who has found that dust attenuation does not play a significant role in the classification of LSBGs when viewed face-on. 

As mentioned in \citet{LEPM22}, we also exclude from this analysis such galaxies with $M_{*} > 10^{12} \Msun$ and $R_{50} >$ 30 kpc. The 30 kpc threshold is a typical boundary employed to separate the Intra Cluster Light (ICL) contribution from the main galaxy \citep{Schaye15, Pillepich18, Henden20,Montenegro23,Montenegro25}, so that we discard those galaxies whose properties could be affected by the inclusion of the ICL component. 11 out of 15 of these extended and massive galaxies in our sample that satisfy such conditions are classified as LSBGs, and are studied in more detail in \citet{Zhu23}. Our final sample consists of 22,539 galaxies, including 5,807 LSBGs and 16,732 HSBGs. Throughout this manuscript, we will focus only on those galaxies labeled as `central' galaxies, which are defined by the \texttt{SUBFIND} algorithm as the most massive galaxy of their parent FoF group. This is due to the fact that most of the properties studied here are well-defined only for central galaxies, especially those associated with their parent dark-matter halo. Given that environmental processes such as ram-pressure stripping, strangulation, and tidal interactions predominantly affect satellite galaxies, while central galaxies are largely governed by the properties of their host haloes \citep{vandenBosch08,Peng10}, restricting our analysis to centrals effectively removes the population most strongly influenced by environmental quenching and rapid transformation processes. Moreover, a significant fraction of central galaxies correspond to isolated systems, further biasing the sample toward field-like environments. Therefore, our results should be interpreted as applying specifically to central galaxies, where internal processes and halo properties dominate, and may not capture the full range of environmental effects present in satellite populations. Our final sample includes 12,220 central galaxies, from which 3,231 and 8,989 are classified as LSBGs and HSBGs, respectively.

\subsection{Supplementary data}
\label{sec:suplementary_data}

The merging history of galaxies can be traced in cosmological simulations using the so-called ``merger trees'', which involves tracing the masses off progenitor haloes and the redshifts at which they merge to form larger haloes. In TNG, merger trees are computed with the \texttt{SUBLINK} algorithm \citep{RodGom15}, which employs a methodology similar to that described in \citet{Springel05} and \citet{Boylan09}. For each subhalo, a unique descendant is assigned as an approximation of the hierarchical buildup structure in $\Lambda$CDM cosmogonies. A single subhalo can have many progenitors, but only a single descendant at most, consistent with the hierarchical buildup of structure in the Universe. Once all the descendant connections are made, the first progenitor of each subhalo is defined as the one with the `most massive history' behind it \citep{DeLucia07}. Therefore, the mass history of any particular galaxy or halo can be robustly compared across simulations. The identification of all the subhalo descendants together with its first progenitor determines the merger trees. Each tree is defined as a set of subhaloes that are connected by progenitor/descendant link or by belonging to the same FoF group.

We also employ the catalogs by \citet{RodGom16} which contain information about the assembly of the stellar component of galaxies in TNG100 across the whole simulation time. These provide measurements of the ex-situ stellar mass fraction of each galaxy, defined as the fraction of the stellar mass of a given galaxy that is formed in a different galaxy at and later accreted through major and minor mergers and flyby encounters, allowing a determination of the `accretion origin' of every stellar particle in the simulation.

\section{Results}
\label{sec:Results}
 
Given that most of the properties studied here are strongly correlated with stellar mass, in order to avoid any bias that could lead into a mis-interpretation of our results our central galaxies are segregated in three different stellar mass ranges centered at 10$^9$, $10^{10}$ and $10^{11} \Msun$ (each of them being a factor of $\sim$2 wide), which throughout this manuscript will be refereed as `low-mass', `intermediate-mass' and `high-mass' galaxies, respectively. In this section, we keep the following convention for subsequent figures with similar format: red and black solid lines show the median evolutionary tracks of LSBGs, HSBGs respectively, while blue solid line shows the median evolutionary track of the ``full'' central galaxy population, i.e., with no surface brightness distinction. The shaded areas enclose the interval between the 16th and 84th percentiles of the given distributions. Errorbars represent the dispersion of the median obtained from a bootstrap re-sampling algorithm, in which a thousand random realizations derived from the original dataset were performed. Dashed vertical lines at $z=5,2$ and $1$ are displayed for visual support of further analysis.

\subsection{Physical parameters of LSBGs and HSBGs progenitors across the cosmic time.}   
\label{sec:GalaxyEvol}
\subsubsection{Galaxy morphology}
\label{sec:Morph}

    \begin{figure} 
        \includegraphics[width=0.47\textwidth]{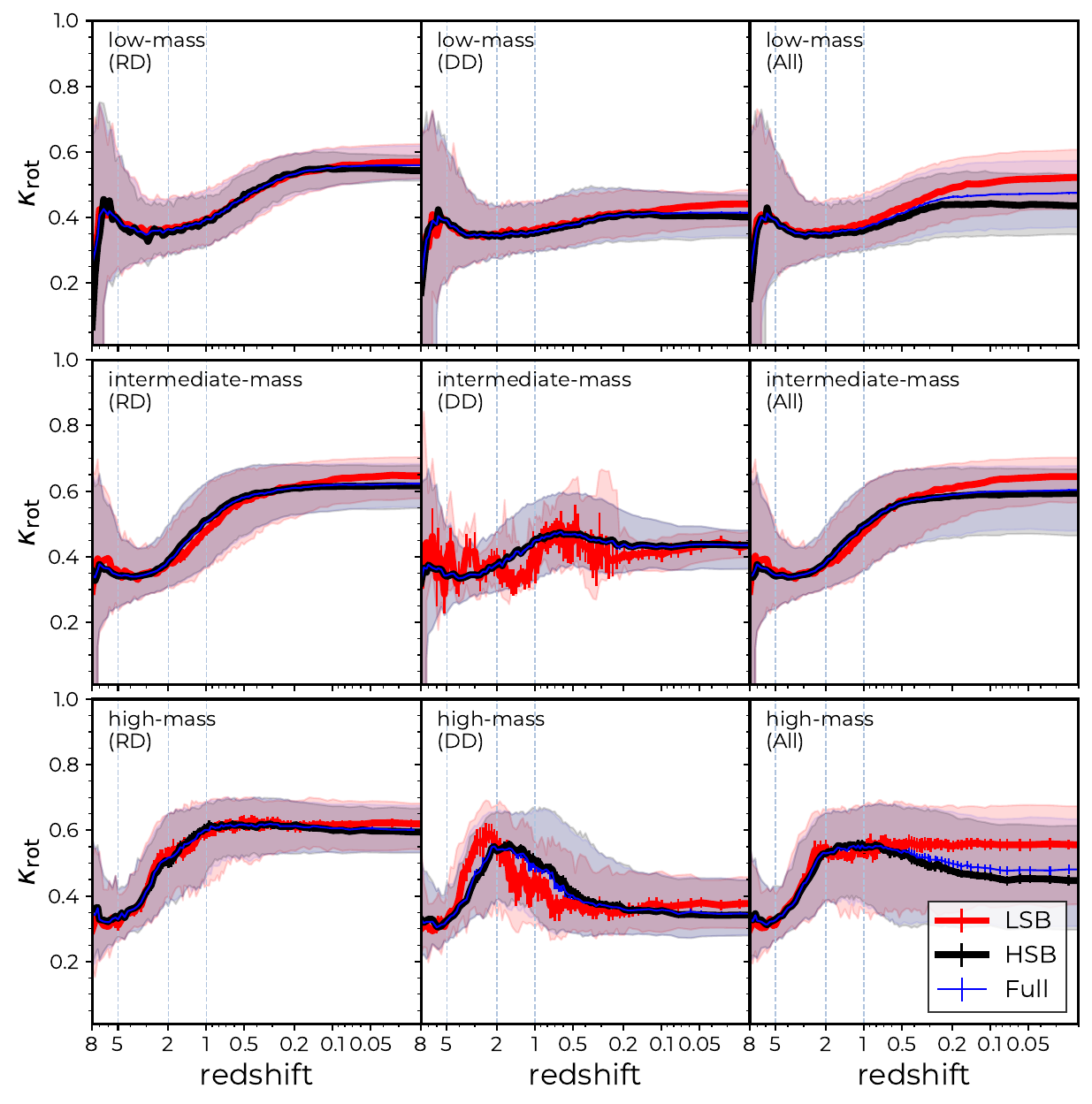} \\
        \caption{Evolution of galaxy morphology for galaxies classified as LSBGs (red) and HSBGs (black) at $z=0$. Rows from top to bottom correspond to three different $z=0$ stellar mass bins centered at $10^9, 10^{10}$ and $10^{11} \Msun$. Errorbars represent the dispersion around the median obtained from a bootstrap re-sampling algorithm, while the shaded regions enclose the interval between the 16th and 84th percentiles. Left and middle columns include galaxies classified as `rotation dominated' (RD) and `dispersion dominated' (DD) at $z=0$, while right column includes all galaxies on a given stellar mass bin, regardless their morphology. Hereafter we keep the same convention for all the figures throughout this manuscript.}
        \label{fig:kappa}
    \end{figure}

In this study, we employ a kinematic criterion (as in previous studies of this series, \citealt{LEPM22, LEPM24}) to estimate the morphology of the galaxies included in the sample, distinguishing between rotation-dominated and dispersion-dominated galaxies. To do so, we implement the definition given by \cite{Sales10}

    \begin{equation}
    \label{eq:kappa}
	       \krot = \frac{K_{rot}}{K},
    \end{equation}
    
\noindent as the ratio between the rotational energy from the azimuthal component $K_{rot}$ of their stellar velocities and the total kinetic energy $K$, so that `rotation-dominated' (RD) galaxies in our sample are those with $\krot > 0.5$, while `dispersion-dominated' (DD) correspond to galaxies with $\krot < 0.5$ at $z=0$. These values are taken from the post-processed catalogs of \citet{RodGom17} at different snapshots. Table \ref{tab:galaxy_sample} summarizes the central galaxies in each stellar mass range of our main sample, classified by their $z = 0$ morphology and surface brightness. As previously highlighted in \citet{LEPM22}, at $z=0$ LSBGs in our sample are mostly rotation-dominated galaxies within a stellar mass range of $9 \lesssim$ log($M_{*}/\Msun$) $\lesssim 11$. We note that this division does not correspond to a clear bimodality in the $\kappa_{\rm rot}$ distribution, but is instead adopted as a convenient operational classification to separate systems with different dynamical support.

\begin{table}
    \centering
    \begin{tabular}{c c c c c c}
        $\Mstar$ & Galaxy &  Rotation & Dispersion & All \\
                 & type & dominated & dominated & morphologies \\
        \hline
        Low & LSBGs &   457 & 291 & 748 \\
            & HSBGs & 248 & 728 & 976 \\
        \hline
        Inter. & LSBGs & 263 & 12 & 275 \\
               & HSBGs & 1056 & 299 & 1355 \\
        \hline
        High & LSBGs & 68 & 36 &  104 \\
             & HSBGs & 138 & 206 &  344 \\
        
        \hline 
    \end{tabular}
    \caption{Number of central galaxies per stellar mass range in our galaxy sample, categorized by their $z = 0$ morphology (rotation-dominated or dispersion-dominated) and galaxy type (LSBGs or HSBGs).}
    \label{tab:galaxy_sample}
\end{table}

Figure \ref{fig:kappa} shows the median evolutionary tracks of $\krot$ The left column corresponds to those galaxies classified as RD at $z=0$,  the middle column corresponding to galaxies classified as DD, and the right column showing the median evolutionary tracks of LSBGs and HSBGs regardless their final morphology. 
    
From the right column, the observed trends indicate that, at recent epochs, the LSBG population is mostly dominated by RD galaxies, regardless of their final stellar mass. In the case of LSBGs, all the evolutionary tracks show a steady state with $\krot > 0.5$ and show almost no variation below a given redshift, whose value increases with stellar mass. A similar behavior is seen in low- and intermediate-mass HSBGs, though with lower $\krot$ values than those of LSBGs. However, massive DD HSBGs are found to reach a maximum in $\krot$, and slowly decreasing with $z$ indicating a significant kinematic evolution at $z<1$, in good agreement with recent studies \citep{Mozumdar25}. Finally Fig. \ref{fig:kappa} indicates that LSBGs in our sample have been predominantly rotation-dominated galaxies since $z\sim1$, with a median value around 0.6\footnote{As previously highlighted in \citet{LEPM22}, around $75\%$ of LSBGs in our sample exhibit morphologies consistent with $\krot > 0.5$ at $z < 1$.}.
    
When morphological segregation is considered, from the left column we can observe that both LSBGs and HSBGs experience a smooth monotonic increase in the values of $\krot$, where galaxy progenitors are transformed from DD ($\krot \sim 0.3$) to RD galaxies ($\krot \sim 0.7$). Interestingly, we found a trend in which the lower the final stellar mass, the lower the redshift in which progenitors experience their transition to DD to RD galaxies. Such transitions occur at $z\sim0.4, 1$ and $2$ for low, intermediate and high-mass galaxies progenitors respectively.

From the middle column, we observe that the morphological evolution of the progenitors of LSBGs and HSBGs exhibits less smooth behaviour. In particular, LSBGs in the central panel of this column (corresponding to intermediate-mass DD galaxies) show noticeably noisy trends. This is primarily due to the very limited number of objects in this subsample (only 12 galaxies; see Table \ref{tab:galaxy_sample}), which amplifies statistical fluctuations. This is reflected in the relatively large error bars shown in the figure, which correspond to bootstrap uncertainties on the median. We note that this increased level of noise is not unique to this panel, but is consistently observed in all figures that involve this specific subsample throughout the present work. In contrast, for high-mass DD galaxies, we observe a smoother trend, with an increase in $\krot$, reaching a maximum at $z\sim2$ and then decreasing. This can be caused by the action of dry minor mergers \citep{RodGom17,Lagos18,Slob25} capable to decrease the angular momentum and disturb the ordered rotation of galaxies. This later scenario is beyond the scope of the current manuscript, but will be discussed in more detail as part of an upcoming work (Pérez-Montaño et al. 2026, in prep.). We finally observe that the maximum value of $\krot$ for LSBGs is reached at slightly higher values of $z$ when compared with HSBGs. After reaching this maximum, the median value of $\krot$ for LSBGs is systematically lower than for HSBGs, and then this trend is inverted at $z\sim0.2$, so that high-mass DD LSBGs exhibit slightly higher values of $\krot$ than HSBGs nowadays. 

\subsubsection{Stellar surface density within $R_{\rm half}$}
\label{sec:density}

    \begin{figure} 
        \includegraphics[width=0.47\textwidth]{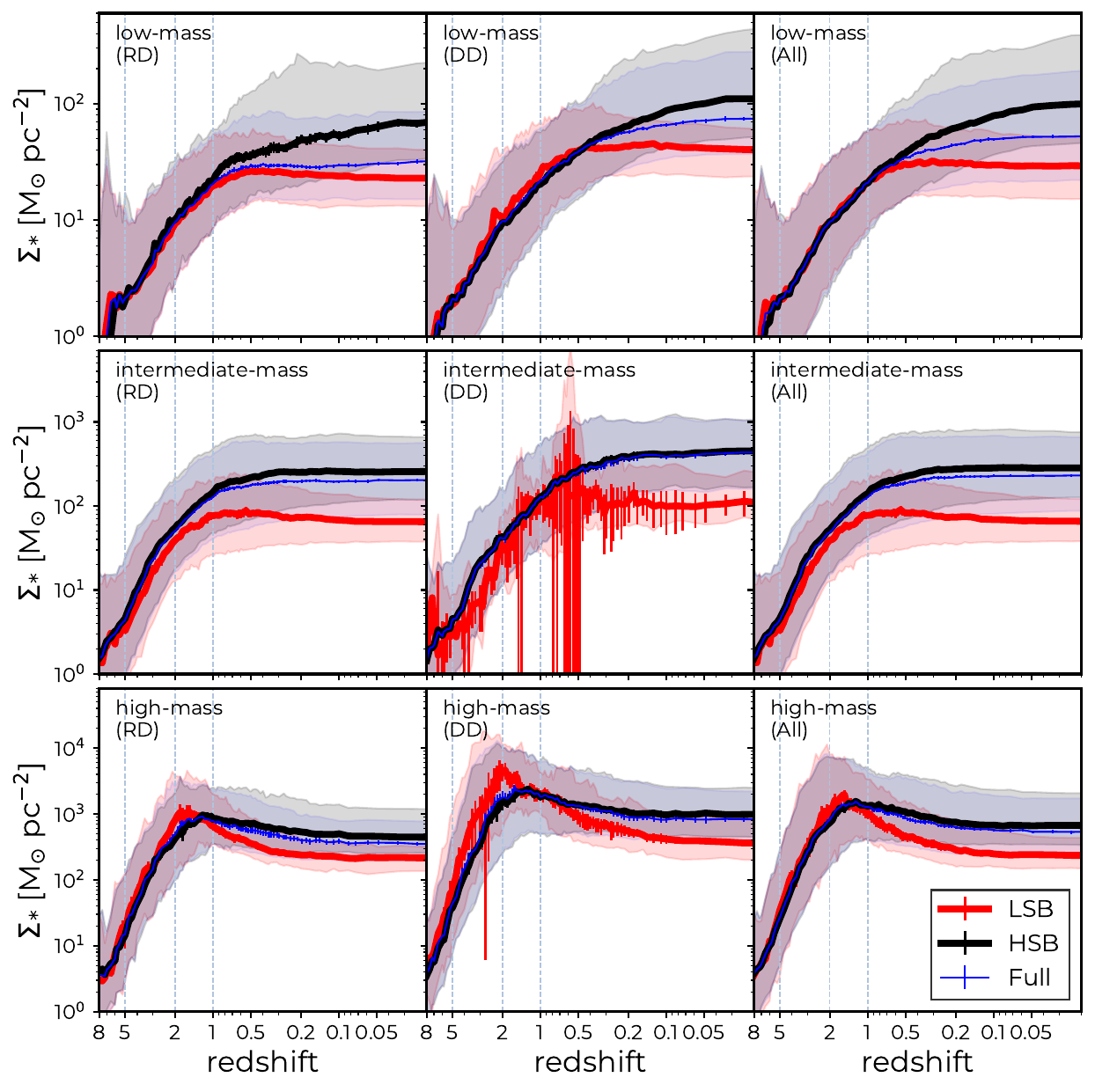} \\
        \caption{Evolution of the central stellar surface density ($\Sigma_*$) defined as in eq. \ref{eq:Sigma}, which is employed as a proxy of the central surface brightness of galaxies in our sample. The median trend indicates that the low-density nature is established around $z=1$, with no significant transitions.}
        \label{fig:density}
    \end{figure}

Given the tight correlation between surface brightness and the stellar surface density, we compute the stellar surface mass density, $\Sigma_{*}$, for all snapshots as follows. First, we project the positions of the stellar particles along the $z$-axis following the procedure described in Section~\ref{sec:sample_const}. We then calculate $R_{half}$, defined as the projected radius containing half of the total stellar mass to obtain the enclosed stellar surface density

\begin{equation}
\label{eq:Sigma}
    \Sigma_* = \frac{M_{*}}{2\pi R^2_{half}}. 
\end{equation}

 We then trace the evolution of $\Sigma_*$ for all progenitors in our LSBGs and HSBGs $z=0$ sample. The results are presented in Fig. \ref{fig:density}, which shows $\Sigma_*$ as a function of redshift. 
 
 From the right column, we observe that for the low and intermediate stellar mass bins, regardless the final morphology of the sample (top and middle row, respectively), the surface-density tracks at high $z$ do not exhibit strong differences between LSBG and HSBG progenitors. However, when such differences arise, present-day LSBGs exhibit systematically lower stellar surface densities than HSBGs, as expected from their low–surface-brightness nature. We also note that the growth in stellar density is always monotonic, reaching a maximum and remaining almost constant below $z\sim 1$. For high-mass galaxies, no significant differences are found between RD LSBGs and HSBGs for ($z>2$), given the overlap of the curves and the confidence intervals. However, at early epochs, LSBGs appear to show higher values of $\Sigma_{*}$ compared to HSBGs, reaching a maximum of $\sim 10^{3-4} \Msun$ pc$^{-2}$ at $z\approx2$, and then slowly decrease inverting the initial trend up to a nearly constant value. 

 Turning our attention to the sample segregated by morphology (left and central columns), the evolutionary tracks are qualitatively similar to the median trend exhibited in the rightmost column. In all cases DD galaxies have higher surface densities than RD ones.

\subsubsection{Specific Angular Momentum}
\label{sec:retention}

    \begin{figure} 
        \includegraphics[width=0.47\textwidth]{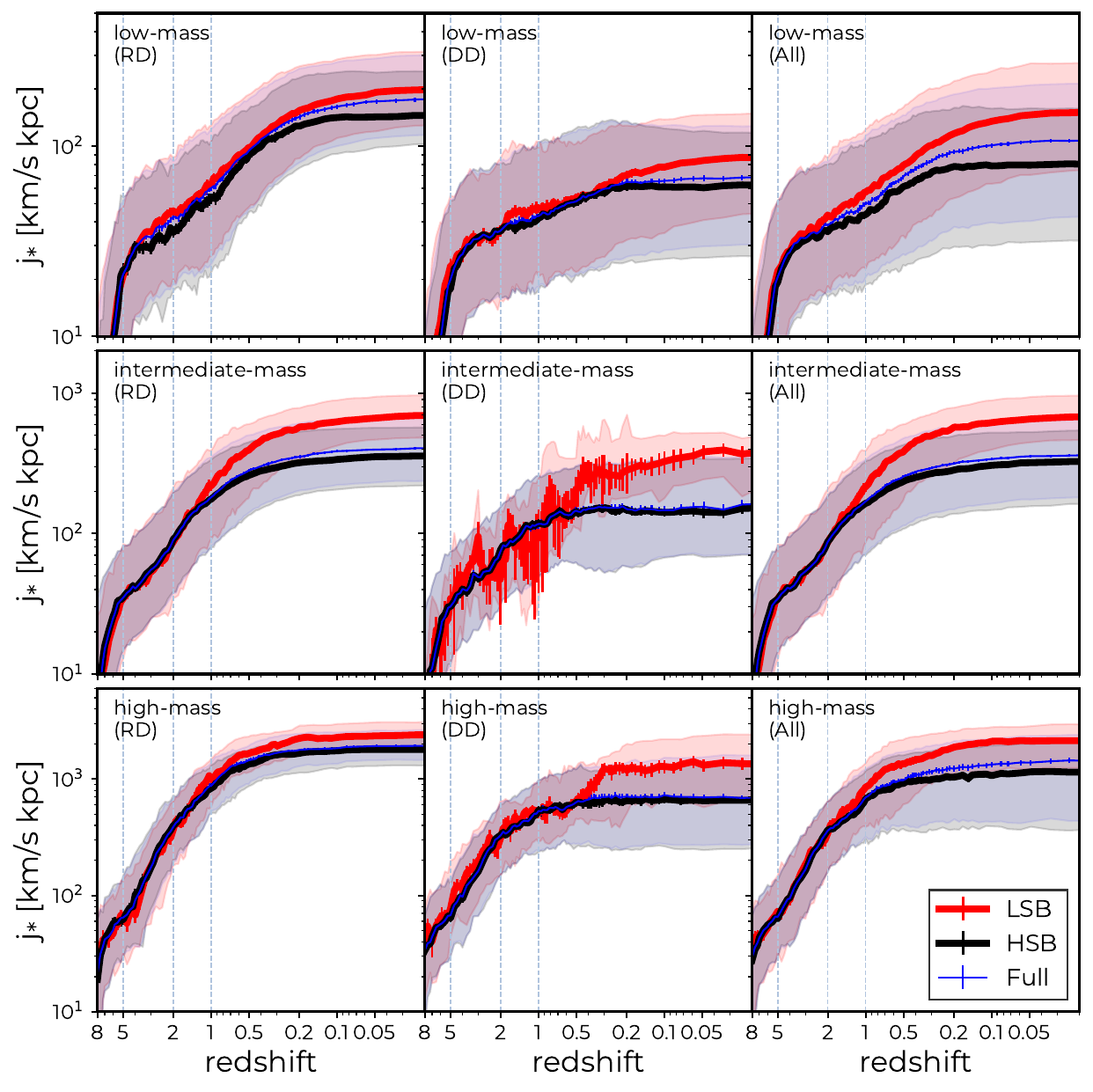}
        \caption{Median evolutionary tracks of the stellar specific angular momentum of LSBGs and HSBGs. In most of the cases, at $z\sim1.5-0.5$, LSBGs progenitors acquire higher amounts of stellar angular momentum than HSBGs progenitors. The divergence in $\jstar$ evolutionary tracks is such that $z_{\Sigma_*} \lesssim z_{\jstar}$, implying a possible cause-consequence relationship between the variations in the angular momentum and the emergence of the low surface density nature.}
        \label{fig:jstar}
    \end{figure}

In order to investigate the origin of the bifurcation in the evolutionary tracks that give rise to LSBGs and HSBGs in our sample, where variations in surface density (and consequently in surface brightness; \citealt{Jimenez98,MMW98}) are expected to be driven by their high specific angular momentum, we show in Fig.~\ref{fig:jstar} the evolution of $\jstar$, computed as

\begin{equation}
\label{eq:angmom}
	\vec{j_{*}} = \frac{\vec{J}}{M} = \frac{\sum_i m_i \vec{r_i} \times \vec{v_i}}{\sum_i m_i},
\end{equation}

where $m_i, \vec{r_i}$ and $\vec{v_i}$ are the mass, position and velocity of a given stellar particle, respectively. In general, we can observe that both LSBGs and HSBGs progenitors constantly acquire angular momentum across their lifetimes, so that the higher the final stellar mass, the higher the final angular momentum, as expected from the existing correlation between $\Mstar$ and $\jstar$ \citep{Fall83,Fall13}.

The left column of Fig. \ref{fig:jstar} shows that, although the evolution of $\jstar$ in both LSBGs and HSBGs is qualitatively similar, around $z\sim1.5-0.5$ LSBGs acquire angular momentum faster than HSBGs, which is reflected in the divergence exhibited in the evolutionary tracks of $\jstar$ between LSBGs and HSBGs. This divergence in $\jstar$ seems to occur roughly before the emergence of the low-density nature of LSBGs (see Fig. \ref{fig:density}), indicating that such variations in the angular momentum of galaxies play a major role on the determination of the low surface brightness nature of the galaxies in our sample.

Similarly, comparing the left-hand and middle panels of Fig. \ref{fig:jstar}, we observe that at fixed stellar mass RD galaxies acquire more angular momentum than DD galaxies, which is also expected from the fact that rotationally supported galaxies have higher angular momentum than those dispersion dominated systems \citep{RomFall12,Fall13,Posti18,DiTeodoro21}. 

    \begin{figure*} 
        \centering
        \begin{tabular}{cc}
        \includegraphics[width=0.47\textwidth]{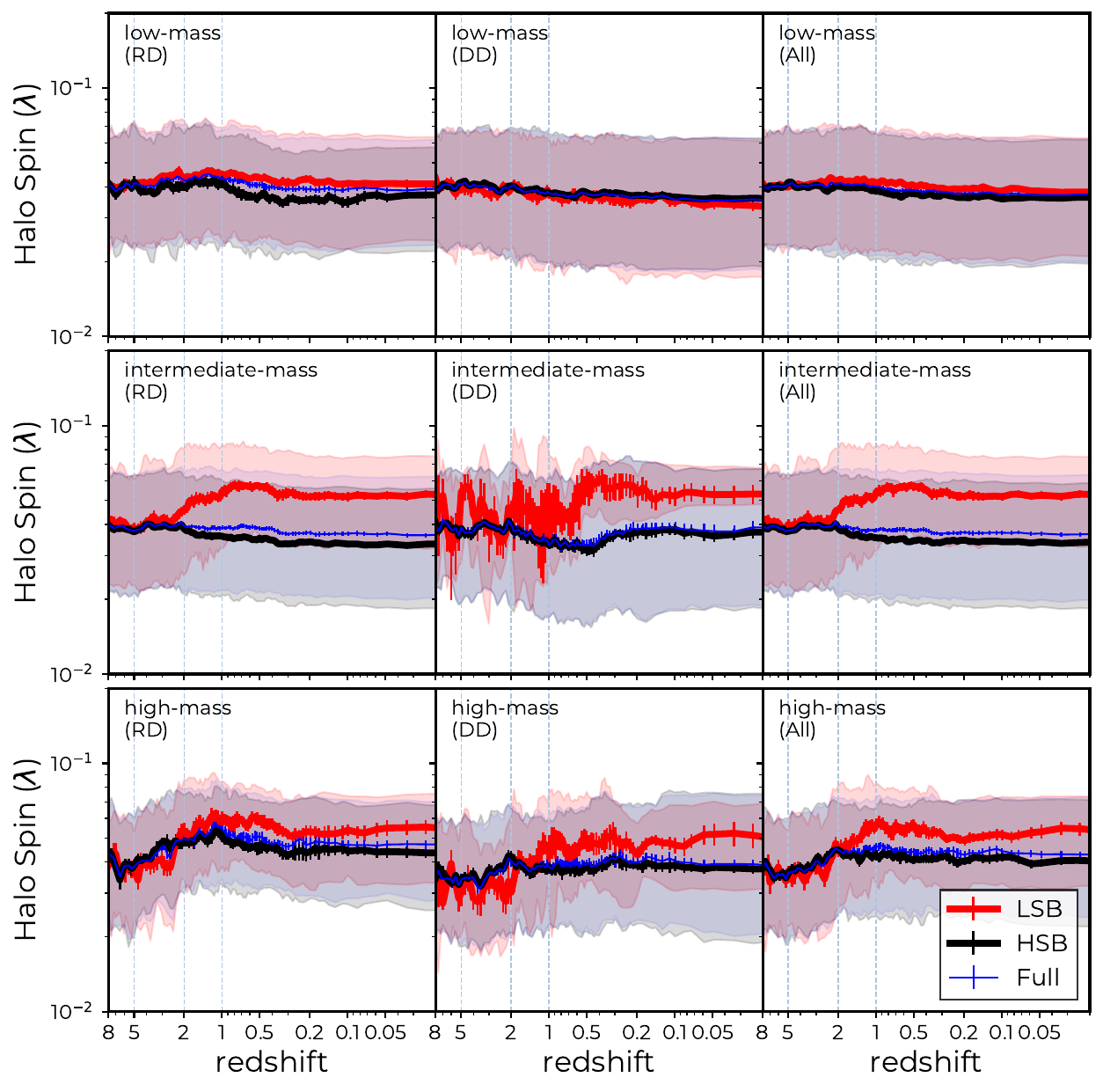} &
        \includegraphics[width=0.47\textwidth]{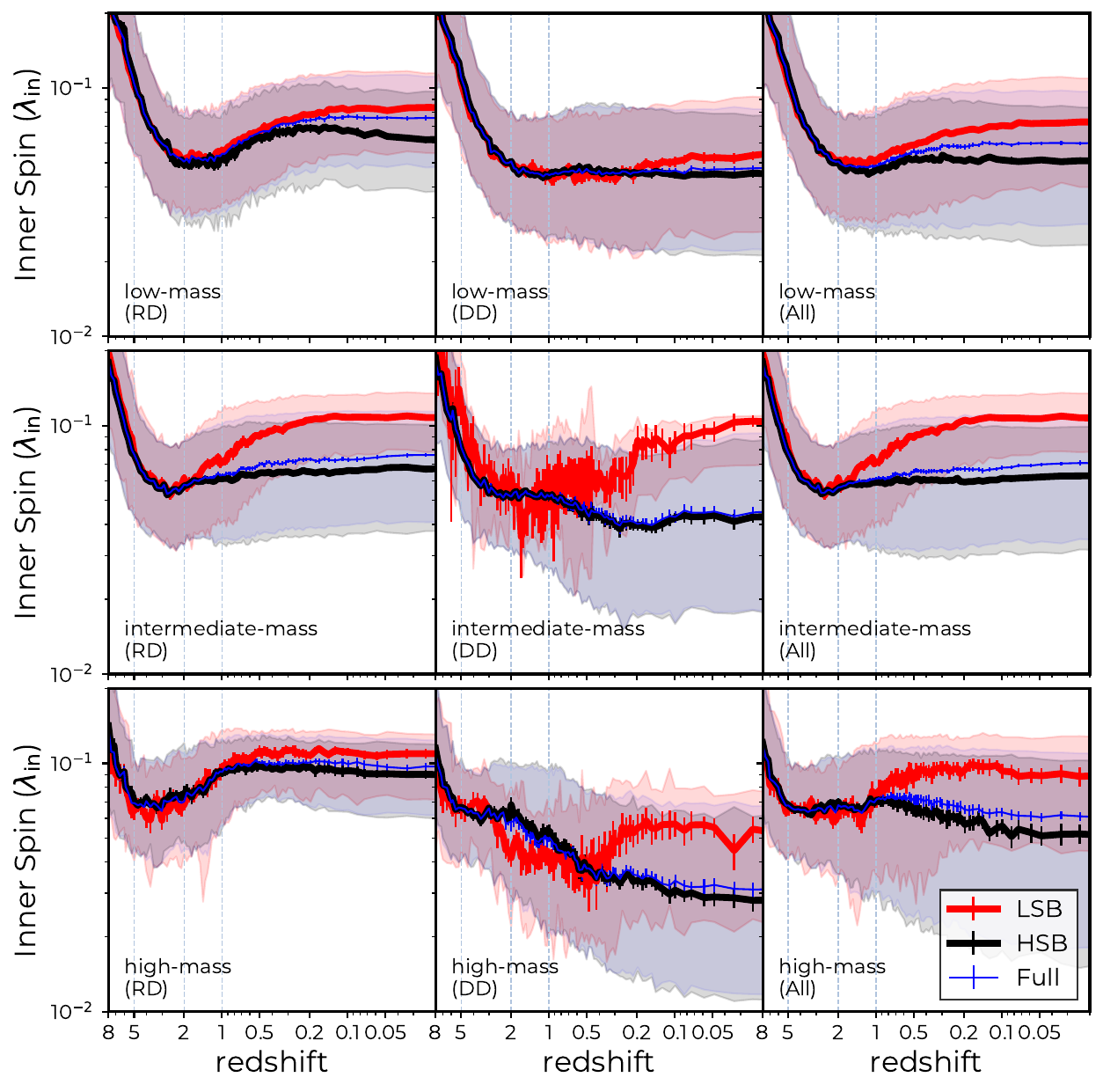} \\
        \end{tabular}
        \caption{\textit{Left Panel:} Spin Parameter evolution of the dark matter halos hosting LSBGs and HSBGs progenitors, computed according to eq. \ref{eq:spin_bullock} and including all the components of the galaxy configuration within $R_{200}$. \textit{Right Panel:} Evolution of the `inner' spin parameter ($\lambda_{in}$), computed within a radius equal to 10\% of $R_{200}$. This quantity has been found to be more closely connected to the stellar morphology of a galaxy \citep{Zavala16} than $\lambda$. Except for massive DD galaxies, in most of the cases $\lambda_{in}$ is found to be higher than $\lambda$ at all redshifts by a factor of $\sim2-3$. }
        \label{fig:spin}
    \end{figure*}

\subsubsection{Halo Spin parameter}
\label{sec:spin_param}

We continue our analysis following the evolution of the halo spin parameter $\lambda$, which is computed following \citet{Bullock01} as

\begin{equation}
\label{eq:spin_bullock}
	\lambda = \frac{j_{200}}{\sqrt{2} R_{200} v_{200}},
\end{equation}

where $j_{200}$ is the specific angular momentum of all the baryonic and dark components within $R_{200}$, defined as the radius of a sphere whose density is 200 times the critical density of the Universe at that time, and $v_{200}$ the circular velocity at $R=R_{200}$. Eq. \ref{eq:spin_bullock} is employed as a proxy of the definition of $\lambda$ given by \citet{Peebles71}. 

The left panel of Fig. \ref{fig:spin} shows the evolution of the spin parameter of LSBGs and HSBGs, in which we can observe a general trend in which the intermediate and high-mass galaxy progenitors follow similar evolutionary tracks at high redshift, and then diverge at $z\approx2$. In all cases, we note that the parent halos of HSBGs exhibit no significant evolution in $\lambda$ throughout their lifetimes, remaining constant at all redshifts. By contrast, LSBGs parent halos exhibit an increase in the median values of $\lambda$ around $z\sim2$ of nearly a factor of 2, and remain almost constant below this redshift. Note that this divergence roughly corresponds to the so-called `turnaround' epoch after which the spin parameter presents no significant changes, consistent with the results of \citet{White84} and \citet{Zavala16}. For low-mass galaxies, the differences between LSBGs and HSBGs become less noticeable, suggesting that external mechanisms, including stochastic processes affecting these systems (which are sensitive to perturbations of any type), may play a more significant role on the emergence of the low density nature of galaxies.

To explore this later hypothesis, the right panel of Fig. \ref{fig:spin} shows the evolution of the inner spin parameter $\lambda_{in}$, defined as the spin parameter of the innermost part of the configuration, computed as in eq. \ref{eq:spin_bullock} but employing only those particles within a radius equal to $0.1R_{200}$. In general the median evolutionary tracks are qualitatively similar to the trends found for the spin parameter of the whole configuration. However, for low mass galaxies the difference in $\lambda_{in}$ between LSBGs and HSBGs seems to be more pronounced when compared to $\lambda$. This result highlights the importance of $\lambda_{in}$ and secular processes in galaxy evolution given its close relation with galaxy morphology as pointed out by \citet{Zavala16}, considering that at scales comparable with $0.1R_{200}$ the galaxy itself contributes the most to the measurement on $\lambda_{in}$. Therefore, for low-mass galaxies we can argue that baryonic processes are more important than the nature of their hosting halos \citep{DiCintio19}.

\subsection{Stellar Assembly}
\label{sec:stellar_assembly}

\subsubsection{Accreted stellar mass fraction}
\label{sec:f_exsitu}

\begin{figure} 
        \includegraphics[width=0.47\textwidth]{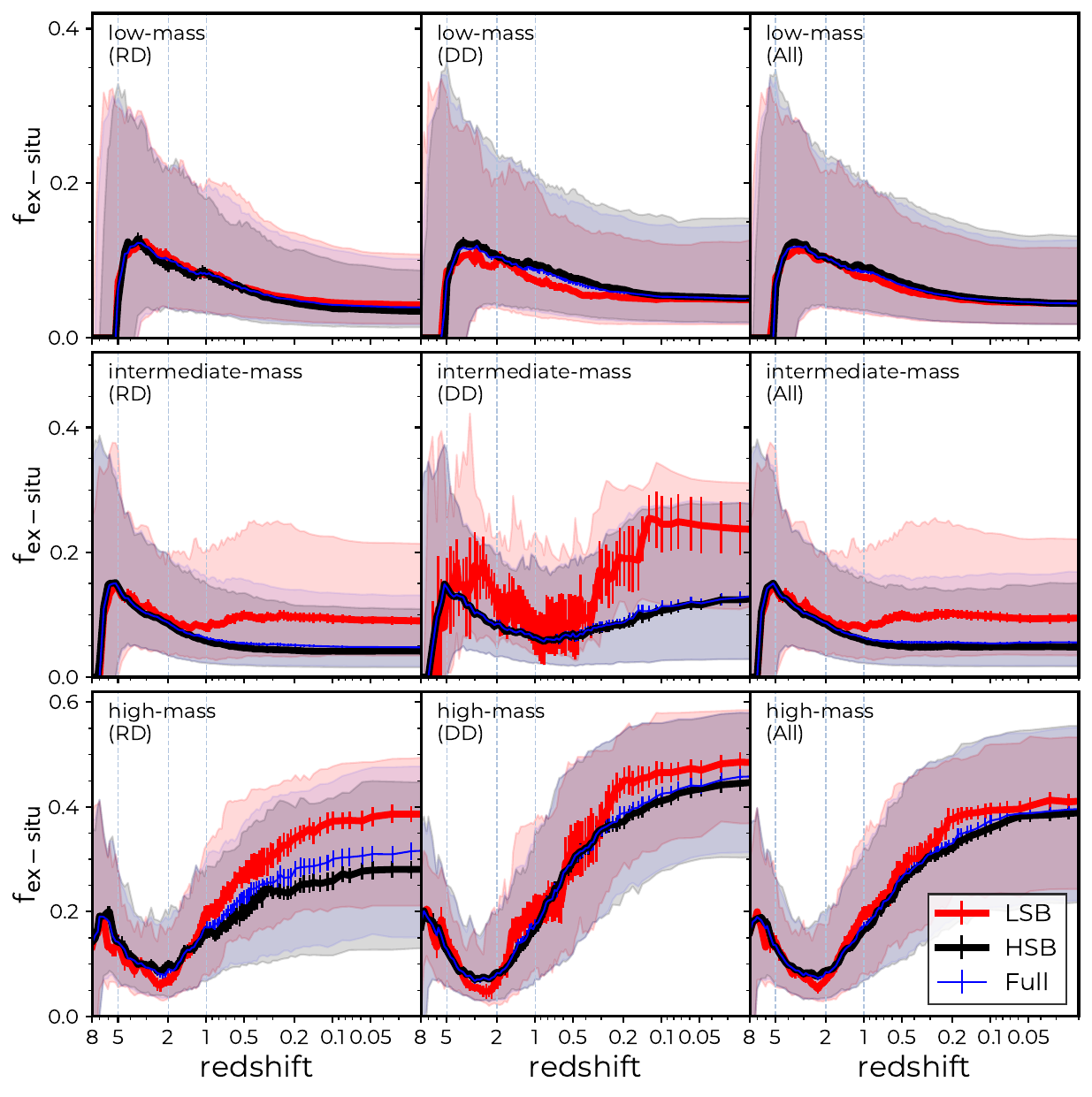} \\
        \caption{The evolution of the accreted stellar mass fraction $f_{ex-situ}$, defined as the ratio between the accreted stellar mass and the total stellar mass of a galaxy at a given redshift. The ex-situ stellar mass consists of the sum of all stellar mass particles that were accreted by complete mergers, ongoing mergers and flybys \citep{RodGom16}. This plot highlights a significant role of galaxy interactions and mergers in the build-up of massive LSBGs, where the ex-situ stellar mass fraction contribute around 40-60\% of the total stellar mass of nowadays LSBGs.}
        \label{fig:ex_situ}
    \end{figure}

Fig. \ref{fig:ex_situ} shows the evolution of the accreted stellar mass fraction $f_{ex-situ}$, defined as the ratio of stellar mass formed `ex-situ', -- i.e. stellar particles not formed within the current subhalo -- to the total stellar mass at a given time. In most cases, LSBGs present higher stellar mass fractions accreted via mergers than HSBGs at $z=0$. As pointed out in \citet{LEPM22}, the `ex-situ' stellar component tends to have a more extended spatial distribution than the `in-situ' stellar component, contributing to the low surface brightness nature of `ex-situ' dominated galaxies, specially the most massive ones.

From the top row of Fig. \ref{fig:ex_situ}, we observe that for low-mass galaxies, the median accreted stellar mass for both LSBGs and HSBGs does not exceed 15\% of the total stellar mass at a given time across the cosmic time. We found no significant difference in the evolutionary tracks between LSBGs and HSBGs. The trends found in the first row could be an indication that mergers do not play an important role on the stellar assembly of low-mass LSBGs, contributing at most 10\% of the total stellar mass at early epochs (with a maximum contribution around 30\%).

In the case of intermediate-mass galaxies (middle row of Fig. \ref{fig:ex_situ}), at $z<5$ LSBG progenitors exhibit systematically higher $f_{ex-situ}$ than HSBG progenitors, regardless their final morphology. For the case of RD galaxies, this difference is more noticeable below $z\sim1.5$, where the median value of $f_{ex-situ}$ in LSBGs is about $\sim10\%$, roughly two times higher than for the case of HSBGs at $z\lesssim1.5$. For DD galaxies, both evolutionary tracks show qualitatively similar trends reaching a maximum at early epochs, then decreasing to a minimum around $z\sim1$, and rising again up to $z=0$. For LSBG progenitors, the increase in $f_{ex-situ}$ is more pronounced than for the case of HSBGs. Although this increase is expected \citep{LEPM22} in both RD and DD galaxies, the noisy trend in $f_{ex-situ}$ observed in DD galaxies (middle row, middle column) can be a consequence of the very low number of DD LSBGs at this stellar mass range (Table \ref{tab:galaxy_sample}), rather than a consequence of a physical process (e.g., major mergers). At $z=0$, the maximum reached after the uprise is $\sim$25\% and $\sim$15\%, respectively. From this row, we can argue that for intermediate-mass galaxies, the role of the accreted stellar component is more or less important depending on the final morphology of galaxies, being more significant in the case of DD LSBGs \citep{RodGom16}.

Finally, in massive galaxies (bottom row of Fig. \ref{fig:ex_situ}) we observe that both LSBGs and HSBGs have similar $f_{ex-situ}$ at early epochs, reaching a minimum at $z\approx2$ and subsequently experiencing an important contribution of accreted stellar mass, so that $f_{ex-situ}$ is continuously increasing with cosmic time at low redshifts, highlighting the role of galaxy mergers in the build-up of massive systems, specially those similar to early-type galaxies, as we can observe in the bottom row of Fig. \ref{fig:ex_situ} where the increase in this quantity is faster in DD galaxies than in RD ones. This effect may be further enhanced by the limited number of objects in some subsamples, which increases the statistical noise and can amplify apparent differences that do not persist in the combined sample.

At $z<2$, LSBGs accrete more stellar mass than HSBGs despite their galaxy morphology, where the bifurcation on $f_{ex-situ}$ between LSBGs and HSBGs occurs at $z\sim1.5$ and $z\sim0.5$ for RD and DD galaxies, respectively. From this, we can argue that mergers and galaxy interactions play a significant role on the development of low-surface brightness features exhibited by massive galaxies, in good agreement with previous studies (e.g., \citealt{Zhu18,Kulier20,Bustos25,Sola25}).

\subsubsection{Stellar Mass Assembly History}
\label{sec:SFhist}

\begin{figure} 
        \includegraphics[width=0.47\textwidth]{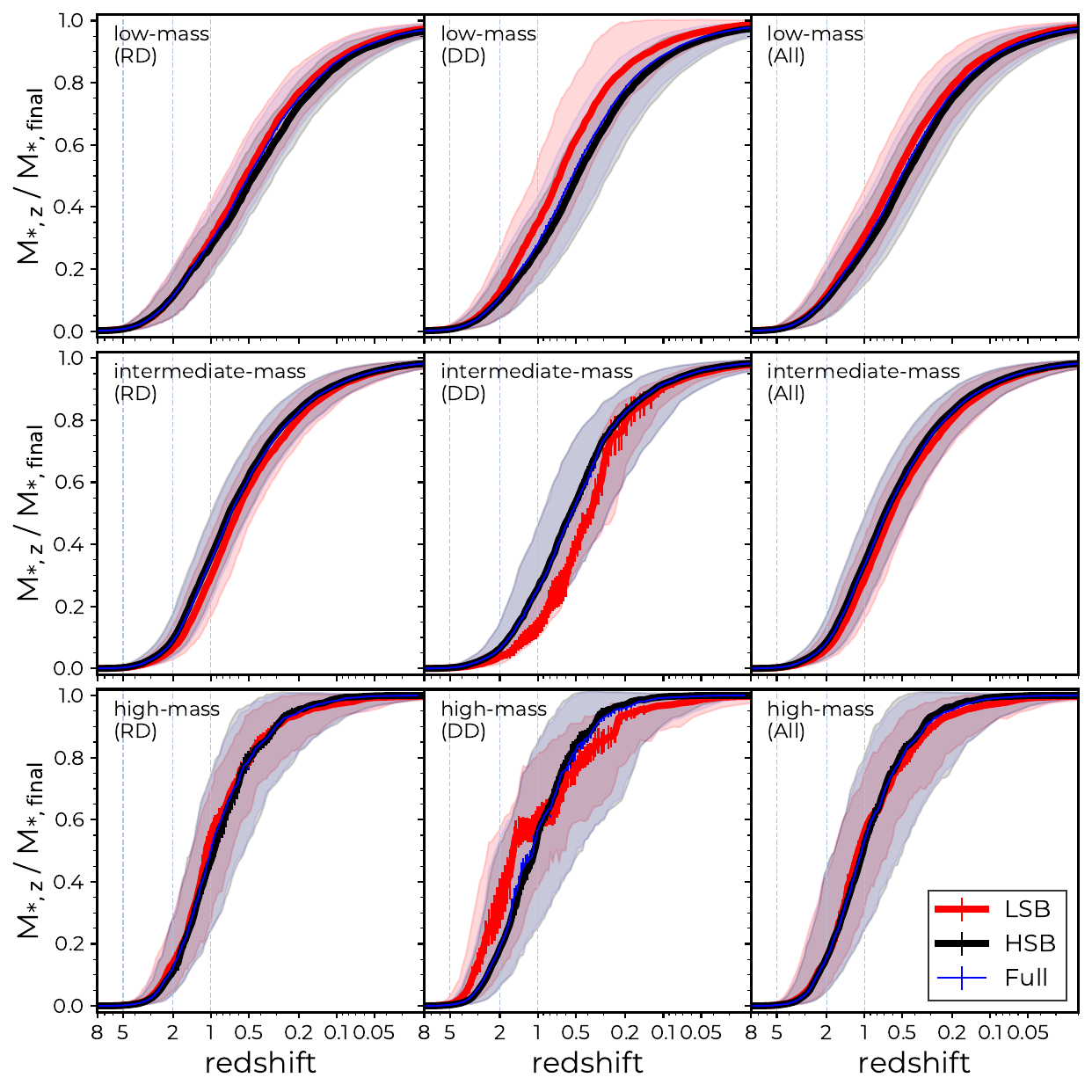} \\
        \caption{Cumulative stellar mass assembly of LSBGs and HSBGs. Rows correspond to different $z = 0$ stellar-mass bins, while columns indicate different morphological types. Overall, the stellar mass assembly histories of LSBGs and HSBGs are very similar, particularly for RD galaxies (left-hand column). Slightly stronger differences are observed for DD galaxies; however, their assembly histories remain qualitatively alike. This suggests that the differences between LSBGs and HSBGs are not primarily driven by ``when'' star formation occurs, but rather by ``where'' new stars are formed within the galaxies.}
        \label{fig:SFH_all}
    \end{figure}

In Fig. \ref{fig:SFH_all}, we plot the median cumulative stellar mass assembly history of the galaxies in our sample. When no morphology segregation is considered (right column), the star formation histories of LSBGs and HSBGs differ only marginally. Although LSBGs have long been suggested to be characterized by low star formation rates \citep{Impey97}, we found that LSBGs and HSBGs of similar stellar mass have similar star formation rates \citep{LEPM22}, in agreement with previous observational \citep{Galaz11,Du15} and theoretical studies employing simulations \citep{Kulier20, Wright21}. This indicates that LSBGs are not inefficient in forming stars in absolute terms, but rather have low star formation efficiencies given their large H I reservoirs \citep{Boissier08,Cao17}.

From the left-hand column of Fig. \ref{fig:SFH_all}, we found that for RD galaxies there is no discernible difference between the stellar mass formation history between LSBGs and HSBGs, suggesting that both galaxy populations may follow similar stellar assembly processes. For DD galaxies (middle-panel of Fig. \ref{fig:SFH_all}) the differences are slightly stronger. We can observe that for the case of low (intermediate)-mass galaxies, LSBGs tend to grow faster (slower) than HSBGs, however qualitatively both LSBGs and HSBGs present very similar assembly histories. Interestingly, for high-mass DD galaxies, HSBGs tend to grow slower until $z\sim1$, and beyond that the trend is inverted, so that HSBGs  evolve faster. These results are consistent with previous observational studies \citep{Boissier03,Schombert13,Schombert14}, as well as those employing simulations \citep{LEPM22,Stoppacher25,Wright25}. Results from Fig. \ref{fig:SFH_all} suggest that the differences between the star formation histories of LSBGs and HSBGs are not related to when stars form, but rather to where they form within the galaxies.

    \begin{figure*} 
        \centering
        \begin{tabular}{cc}
            \includegraphics[width=0.47\textwidth]{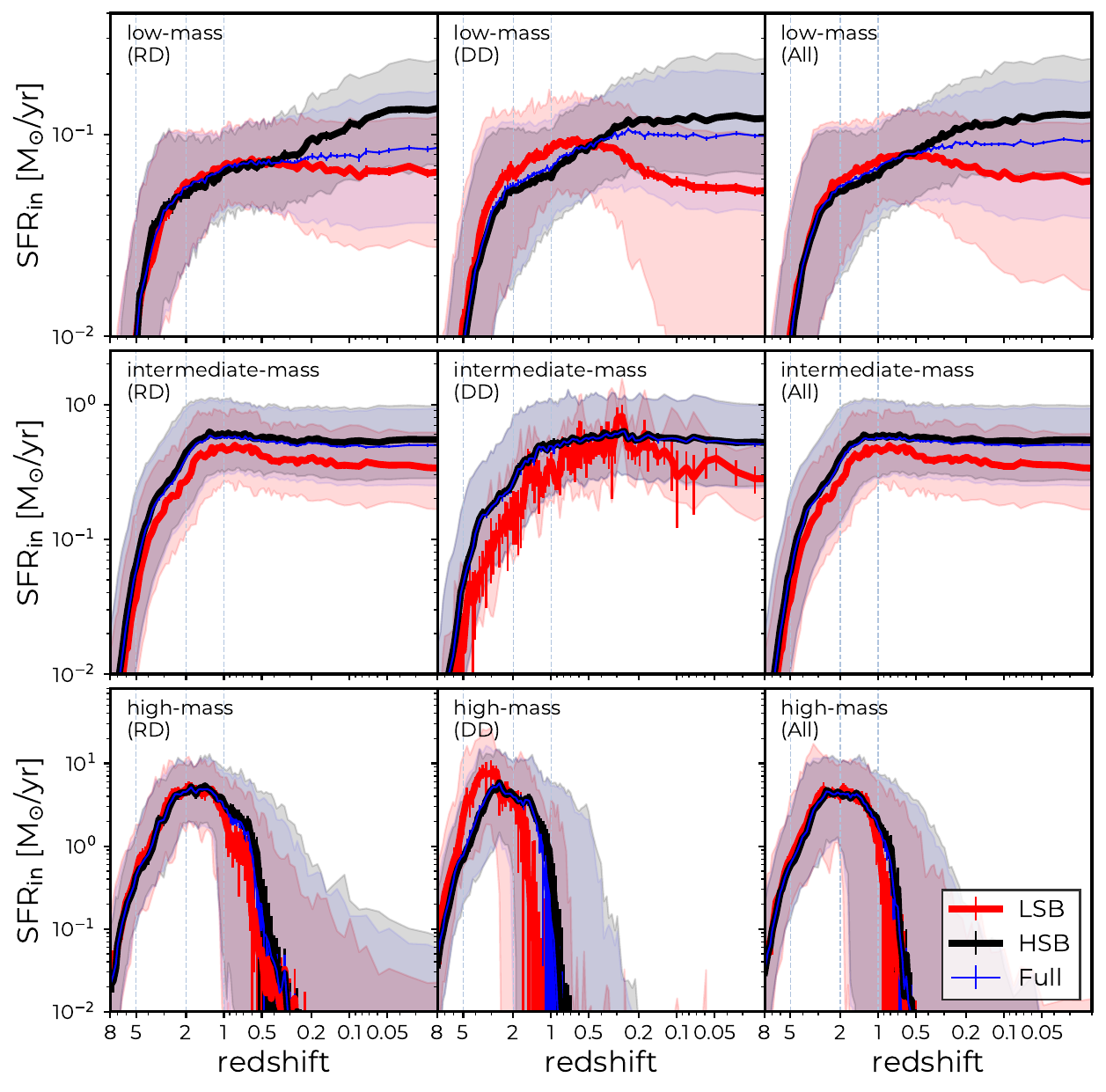} & 
            \includegraphics[width=0.47\textwidth]{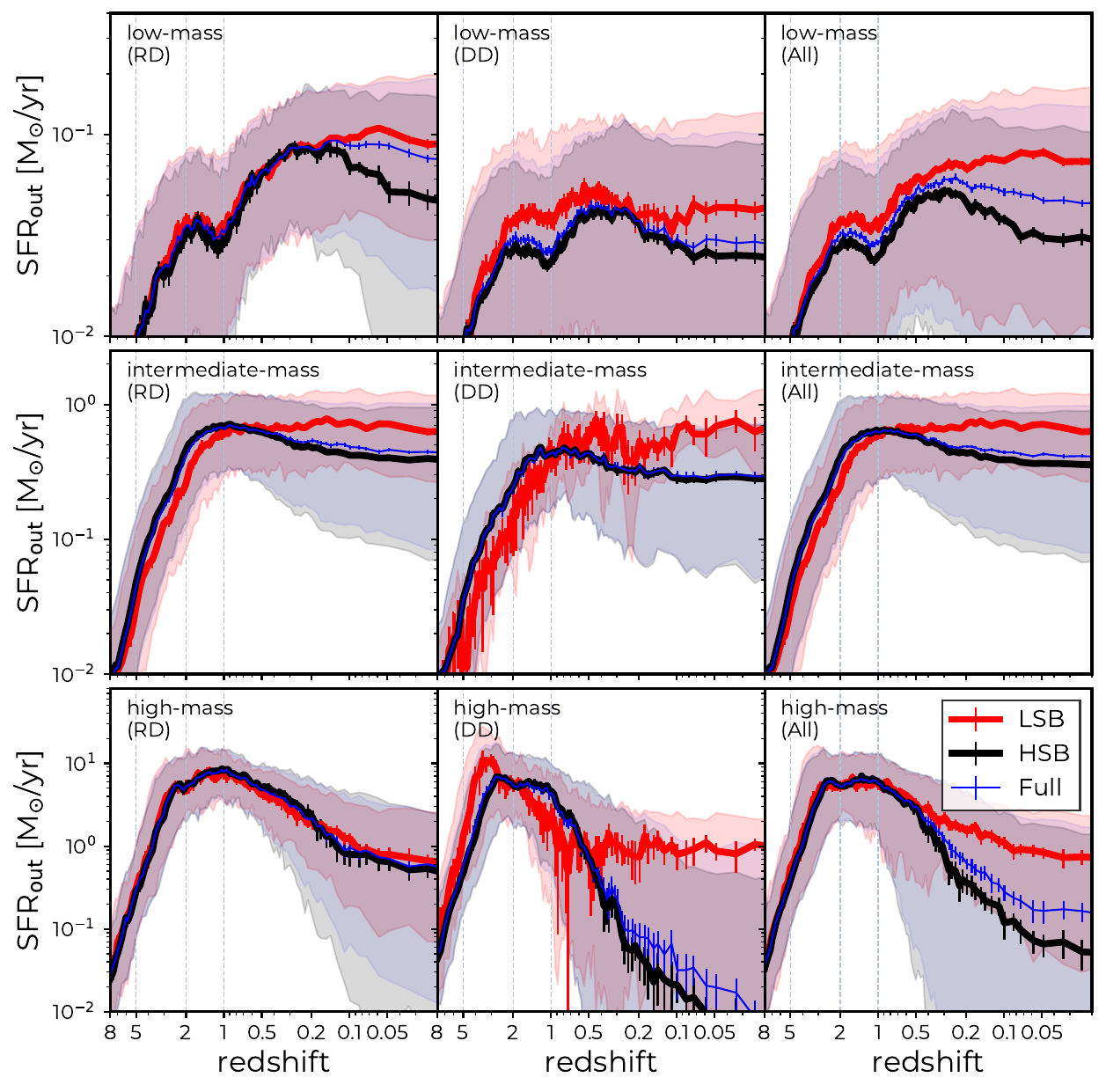} \\
        \end{tabular}
        \caption{Star formation histories of LSBGs and HSBGs within the effective radius (left-hand panel) and beyond it (right-hand panel), which here after are labeled as the `inner' and `outer' SFR. Rows correspond to our different $z=0$ stellar mass bins while columns indicate different morphological types. In general cases, LSBGs exhibit higher SFR in their outskirts when compared to HSBGs, specially at recent epochs ($z < 0.5$), with marked differences on their evolutionary tracks which depend on both, galaxy morphology and stellar mass.}
        \label{fig:SFH}
    \end{figure*}

To test this hypothesis, we continue our analysis by exploring the star formation histories of galaxies in our sample. Left-hand panel of Fig. \ref{fig:SFH} shows the evolution of the `inner' SFR (SFR$_{in}$), defined as the SFR within the stellar half mass radius, while right-hand panel shows the evolution of the `outer' SFR (SFR$_{out}$), defined as the SFR beyond the stellar half mass radius. We can observe that both the inner and outer star formation rates exhibit different evolutionary tracks depending on their morphology and stellar mass range. Nevertheless, a common trend emerges from both panels: LSBGs exhibit higher SFR in their outskirts when compared to HSBGs, especially at recent epochs ($z > 0.5$), which is consistent with previous studies of LSBGs employing simulations, highlighting the strong star formation activity in galactic outskirts \citep{Zhu18,DiCintio19,LEPM22}. This can be caused by a combination of both, the accretion of high angular momentum material from gas-rich galaxies, as well as an increase in halo spin parameter in LSBGs that redistributes star formation, reducing their central star-formation activity and pushing it to the outskirts \citep{Lagos18, Wright21, Wright25}.

Another interesting feature observed in the right-hand panel of Fig. \ref{fig:SFH} is that, in most cases, there is a bifurcation in the outer SFR that roughly coincides with a higher accreted stellar mass fraction as indicated in Fig. \ref{fig:ex_situ}. While the enhanced outer SFR observed in LSBGs is broadly consistent with their higher accreted stellar mass at recent epochs, the detailed evolutionary tracks depend on stellar mass and morphology, as discussed below.

In the low-mass regime (Fig. \ref{fig:SFH}, top row) LSBGs exhibit systematically higher outer star formation rates than HSBGs at all epochs. At early epochs, LSBGs exhibit higher SFR$_{in}$ when compared to HSBGs, then at $z=0.5$ the trend is inverted, such as LSBGs seem to be more quenched in their inner regions when compared to their HSBGs counterparts. At low redshift ($z<0.5$), RD LSBGs show nearly constant SFR$_{in}$, while DD LSBGs display a declining trend, consistent with reduced central star formation. Overall, these trends are suggestive of an `inside-out' star formation scenario, in which older stellar populations dominate the inner regions.

In the intermediate-mass regime (Fig. \ref{fig:SFH}, middle row), the trends are largely reversed with respect to low-mass galaxies, that is, the SFR$_{in}$ in LSBGs is systematically lower than in HSBGs at low redshifts. Additionally, the evolution of SFR$_{out}$ follows a trend in which at early epochs, LSBGs have lower star formation at their outskirts compared with HSBGs up to $z\sim1$. Below this redshift, such trend is inverted, so that SFR$_{out}$ in LSBGs is enhanced with respect to HSBGs regardless their morphology, ending with nowadays LSBGs that are constantly forming stars in their outer regions. In both DD and RD galaxies, the inner and outer SFR exhibit a similar behavior in which the SFR increase from early epochs, reaching a maximum and then remaining almost constant below $z<0.5$.

Finally, in the high-mass galaxy population (Fig. \ref{fig:SFH}, bottom row) we observe that both the inner and outer SFR are quite similar at early times. However, below $z\sim0.5$ LSBGs and HSBGs show a quenched SFR$_{in}$. At $z\lesssim0.5$ all the star formation processes occur mostly in the galactic outskirts. In the case of high-mass RD galaxies, the SFR$_{out}$ remain nearly constant, with only a mild decline toward low redshift. For massive DD galaxies, we note that LSBGs exhibit very high SFR$_{out}$ with respect to HSBGs, indicating that the former are continuously forming new stars, where the outer SFR remains almost constant below $z=0.5$. In contrast, below $z=0.5$ the outer SFR in HSBGs is continuously decreasing, which is an indication of quenching in DD HSBGs (e.g. \citealt{Slob25}).

\section{Discussion }
\label{sec:Discussion}

The results presented in Section \ref{sec:Results} indicate that the evolutionary pathway of LSBGs is primarily driven by variations in their angular momentum distribution and halo spin parameter at  $z \sim 2$, with environmental effects and galaxy interactions playing a secondary role as demonstrated in \citet{LEPM24}, influencing other physical properties such as galaxy morphology and size once the LSBG nature is established by their high angular momentum.

\subsection{The emergence of the LSBG nature.}

The emergence of a stable low surface brightness nature has been explored in previous studies using simulations. In \citet{DiCintio19} the authors highlight that certain major merger configurations are able to increase (decrease) the angular momentum of the primary galaxies, resulting in a lower (higher) surface brightness. Below $z=2.5$ no significant changes in the surface brightness of galaxies are observed once they are classified either as LSBGs or HSBGs. Similarly, \citet{Stoppacher25} found that the values of $z$ at which angular momentum and size evolutionary tracks diverge is such that $z_{size} \lesssim z_{j*}$\footnote{\citet{Stoppacher25} defined $z_{\mathrm{size}}$ and $z_{j_*}$ as the redshifts at which the evolutionary tracks of galaxy size and specific angular momentum, respectively, diverge between LSBGs and HSBGs.}, which is reflected in the divergence of $\mu_r$ at similar redshifts. Moreover, they also find that LSBGs are found to maintain their nature since $z\sim 2-3$. This is consistent with our findings described in sec. \ref{sec:density}. Although we do not directly explore the changes in the values of $\mu_r$ in our sample, we can approximate this to the evolution of their stellar mass surface density in Fig. \ref{fig:density}. In all cases we observe no significant changes in $\Sigma_{*}$ below $z \approx 2$, which is consistent with the $z_{size} \lesssim z_{j*}$ trends found in \citet{LEPM22} and \citet{Stoppacher25}, indicating that the low central density (and consequently, their low surface brightness nature) is kept below $z\sim 2$.

\subsubsection{Variations in Galaxy Morphology.}

In Fig. \ref{fig:kappa}, we find that LSBGs in our simulation follow similar evolutionary trends across all final stellar-mass bins, showing a smooth increase in $\krot$ with time, which indicates that galaxies become more rotationally supported at low redshift. In all three stellar mass bins LSBGs consistently exhibit higher $\krot$ values compared to HSBGs. Comparing with \citet{LEPM22}, we find that $z_{\krot} < z_{j*}$, suggesting that the divergence in the evolutionary tracks of $\krot$ occurs after that of $\jstar$, in agreement with our previous findings. Additionally, in Fig. \ref{fig:density}, we observe that $z_{\krot} < z_{\Sigma_{*}}$, implying that the LSBG nature of galaxies is established before their final $\krot$ values, suggesting that the final morphological state, whether RD or DD, is not a primary driver of the LSBG nature, but rather a consequence of it. We also note that the RD/DD classification is based on a fixed threshold in $\kappa_{\rm rot}$, which may introduce additional scatter when analysing small subsamples.

The most erratic morphological evolution is observed in massive galaxies with nowadays DD morphologies. Both LSBG and HSBG progenitors evolve from lower to higher $\krot$ values between $5 < z < 2$, indicating that massive galaxies undergo significant changes, progressively becoming DD below $z=2$. This behavior aligns with recent observational studies of high-redshift galaxies, such as \citet{Sampaio25}, where it was found that at low stellar masses, the fraction of disc- and bulge-dominated galaxies remains relatively constant at $z \leq 2.4$, while at higher masses, this fraction evolves in opposite directions, suggesting that massive disks are progressively transformed into spheroids. Similarly, \citet{Mozumdar25} observed comparable trends using IFU data for galaxies at $0.25 < z < 0.75$, concluding that massive early-type galaxies have experienced significant kinematic evolution, losing angular momentum as they evolve toward the present days.
    
Moreover, \citet{Sampaio25} identify two galaxy populations, one with diffuse light profiles, suggesting an in-situ formation mechanism which is tied to angular momentum retention during collapse, similar to our LSBGs. The second population has more concentrated, steeper light profiles, indicative of dissipative processes, akin to HSBGs in our sample.

\subsection{The assembly of LSBGs and HSBGs.}
\subsubsection{The role of mergers and accreted stellar mass}

Observations of massive galaxies at different redshifts have shown that, as the Universe ages, galaxies grow in size more than in stellar mass \citep{HuertasCompany13,VanDerWel14}. Most studies suggest that minor mergers explain the continuous size growth of galaxies \citep{Bezanson09,Naab09,VanDerWel09,Matharu19}. However, the role of mergers in building LSBGs appears significant only in massive galaxies. \citet{Naab09} found, using a high-resolution cosmological simulation, that the central part of a massive galaxy is dominated by `in-situ' formed stars, while at radii beyond 2-3 kpc (below $z \approx 2-3$), the system is dominated by accreted stars. Such trend of high-mass DD systems to be more sensitive to mergers was also found by \citet{Stoppacher25}. This is consistent with the last rows of Figs. \ref{fig:ex_situ} and \ref{fig:SFH}. While it remains unclear if minor mergers fully account for size growth, if they do drive the size evolution of massive LSBGs, then a differential size increase would be expected.  

Overall, the differences between galaxy populations can be attributed to gas availability for `in-situ' star formation in RD galaxies and the number of encounters with massive satellites that increase $f_{ex-situ}$ in DD systems. Recent observational studies such as \citet{Bustos25} highlight the role of interactions in the development of massive extended discs, such as Malin 1 \citep{Bothun87}, one of the most iconic LSBGs. The authors found that past interactions contributed to Malin 1's morphological features, such as its gas-rich nature and extended disk. Similar findings were reported by \citet{Zhu18} for a Malin 1 analogue in TNG100, where recent interactions triggered gas cooling, forming an extremely extended stellar disc. \citet{Sola25} used deep optical imaging of low surface brightness structures in massive galaxies from the ATLAS$^{3\mathrm{D}}$ catalog \citep{Capellari11}, finding that the fraction of galaxies with faint, tidally-driven structures increases with stellar mass. \citet{Lee24} observed that luminous galaxies in TNG50, located in dense regions, tend to maintain their size due to angular momentum loss during interactions, while dark matter-dominated galaxies expand due to their isolated nature. Finally, \citet{Pallero25} found that massive disk galaxies in TNG100 generally have a quiescent merging history, with most of their stellar mass formed via star formation. Once these disks are massive enough, they are rarely disrupted due to their presence in low-density environments, similar to our simulated LSBGs \citep{Zhu23,LEPM24}.

Our results in Fig. \ref{fig:ex_situ} show that galactic mergers and tidal interactions, reflected in the accreted stellar mass fraction, play a minimal role in the assembly of low-mass LSBGs, with both galaxy populations having comparable $f_{ex-situ}$ values (median less than 10\%, max $\sim$30\%). For intermediate-mass galaxies, LSBG assembly is also not strongly influenced by mergers, particularly for RD galaxies, though they have higher $f_{ex-situ}$ values than low-mass LSBGs. In contrast, Fig. \ref{fig:ex_situ} emphasizes the role of interactions in high-mass LSBGs assembly, where accreted stellar mass constitutes $\sim$40-60\% of total stellar mass. This is consistent with \citet{RodGom17}, which investigated the joint effect of halo spin and merging history on galaxy morphology, finding that halo spin dominates at low masses while dry mergers are more important for higher-mass systems. \citet{Rodriguez25} also found larger $f_{ex-situ}$ for spheroid-dominated galaxies than for disc-dominated ones with $\Mstar=10^{10-11}\Msun$, with the former having $\sim30\%$ of their stellar mass from mergers, in agreement with our results. 
    
Despite LSBGs being commonly found as dark matter-dominated systems \citep{Pickering97,McGaugh01}, some studies suggest that LSBGs and HSBGs inhabit similar halos at a given stellar mass \citep{Kulier20}. In our sample, halos hosting LSBGs have slightly lower stellar-to-halo mass ratios than HSBGs \citep{LEPM22}, with RD and DD LSBGs dominating low- and high-density regimes, respectively \citep{LEPM24}. Therefore, mechanisms like mergers and galaxy interactions not only increase and redistribute stellar content in LSBGs, but also influence their spin parameter and angular momentum (as shown in Figs. \ref{fig:jstar} and \ref{fig:spin}), contributing to significant changes in morphology, size, and stellar surface density, ultimately leading to the formation of lower surface brightness galaxies. This later result is also in line with \citet{Stoppacher25}, who also found that LSBGs in general have higher $R_{v_{max}}$ at $z \lesssim 1.5$, meaning that LSBGs are in general less concentrated.

\subsubsection{Star Formation Histories of LSBGs and HSBGs}

Fig. \ref{fig:SFH_all} shows that star formation histories of LSBGs and HSBGs are similar (except for massive DD galaxies), as pointed out by several authors \citep{Boissier03,Galaz11,Martin19,LEPM22,Wright25}. We observe that LSBGs exhibit steady star formation rates across cosmic time, driven by their high gas reservoir and low star formation efficiencies \citep{Boissier08,Cao17,Kulier20}, without strong star formation bursts. However, as seen in Fig. \ref{fig:SFH}, the differences in star formation histories become evident when exploring their spatial distributions.

Previous studies \citep{Zhu18,DiCintio19,LEPM22} have shown that accretion of high angular momentum material from gas-rich galaxies enhances star formation at large radii. Similarly, \citet{Wright21,Wright25} proposed that an increase in halo spin parameter in LSBGs redistributes star formation, reducing central activity and pushing it to the outskirts. As a result, low surface brightness galaxies have more extended star formation histories, with most ongoing star formation occurring in their outer regions in line with recent observational studies such as \citet{Shen26}.

In previous studies of this series \citep{LEPM22,LEPM24} we found that merging history has a very small effect on the formation and evolution of LSBGs, at least in a statistical sense. Preliminary results from Pérez-Montaño et al. 2026 (in prep.) suggest changes in the spatial distribution of $\jstar$ due to mergers, rather than changes in the total value of $\jstar$. These results suggest that the radial distribution of the stellar angular momentum becomes more extended in galaxies with higher \textit{ex situ} stellar mass fractions such as LSBGs in our sample, specially those with intermediate and high stellar masses. This angular momentum redistribution seems to be the main reason of the enhanced star formation processes in galactic outskirts.

These figures show that low- to intermediate-mass galaxies sustain star formation over cosmic time, with no clear evidence of recent quenching. In contrast, massive galaxies—both LSBGs and HSBGs—exhibit a rapid increase in their inner star formation rates at early epochs, peaking at $z\sim 2$, followed by central quenching around $z \approx0.5$. However, their outer regions display a markedly different behavior: while massive HSBGs show signs of quenching also in their outskirts, massive LSBGs maintain ongoing star formation in their external regions down to $z=0$. This pattern is consistent with an inside-out quenching scenario for HSBGs, whereas massive LSBGs retain extended star formation histories dominated by their outer discs.

This scenario for massive galaxies aligns with \citet{Naab09}, who found that early mass evolution at $z>2$ is driven by enhanced star formation in the innermost regions, with a decreasing contribution towards $z\approx0.7$. Below this redshift, only a few stars form within 30 kpc, indicating a central stellar density decrease by an order of magnitude at lower redshifts.

\section{Conclusions}
\label{sec:Conclusions}

This is the third part of a series of manuscripts related to the LSBGs population in the TNG100-1 run of the IllustrisTNG project, whose physical properties and environment at $z=0$ has been previously characterized in \citet{LEPM22} and \citet{LEPM24}. The simulated sample includes central galaxies with stellar masses in the range $10^9 \, \Msun < M_{*} < 10^{12} \, \Msun$ classified as LSBGs or HSBGs based on a central surface brightness threshold of $\mu_r = 22.0$ mag arcsec$^{-2}$. 
    
We have used the \texttt{SUBLINK} merger trees \citep{RodGom15} to characterize the main branch of the $z=0$ LSBGs and HSBGs classified as central galaxies. This approach allows us to trace the evolution of their physical properties by linking the current subhalo to its corresponding `first progenitor' at every snapshot, ensuring that the evolution follows the main branch of the current $z=0$ subhalo, as well as the parent halos in which these two galaxy populations reside.
    
To eliminate any potential stellar mass dependence in the interpretation of our results, we segregate central galaxies into three stellar mass ranges: centered at $10^9 \, \Msun$, $10^{10} \, \Msun$, and $10^{11} \, \Msun$. These galaxies are further classified as `rotation-dominated' (RD) or `dispersion-dominated' (DD) at $z=0$, according to eq. \ref{eq:kappa}. The main findings of the present work are as follows:

    \textit{i)} Approximating central surface brightness evolution as that of central stellar density, we observe no significant changes in $\Sigma_*$ below $z\approx 2$, suggesting that the low-density nature of LSBGs is preserved for a significant part of their lifetime. 

    \textit{ii)} Although both galaxy populations undergo multiple transformations over cosmic time, LSBGs smoothly evolve towards higher $\krot$ values compared to HSBGs. Once their low surface brightness nature is established, morphological changes do not affect this characteristic.

    \textit{iii)} Mechanisms such as mergers and galaxy interactions can increase and redistribute stellar content in LSBGs, while also altering or maintaining their spin parameter and angular momentum, leading to changes in size and stellar content. However, mergers seem most relevant for building LSBGs in massive galaxies.

    \textit{iv)} The star formation histories of LSBGs and HSBGs show no significant differences, but differ in their spatial distribution. In LSBGs, ongoing star formation mainly occurs in the outskirts, rather than in the innermost regions.

All together, our results suggest that the mechanisms favoring the low surface brightness nature are closely linked to variations in the spin parameter and angular momentum of host halos. These variations are likely induced by major mergers in the early stages of galaxy evolution, which tend to transfer a fraction of the stellar angular momentum from the inner regions of a galaxy to larger radii. In contrast, stars contributed through minor mergers, as well as those stripped during galaxy flybys, predominantly deposit their angular momentum in the outer regions. As a result, these processes can substantially enhance the stellar angular momentum of the central galaxy at large galactocentric distances, leading to a redistribution of angular momentum, stars and gas from the inner regions to the outskirts, leading to a decrease in central surface brightness. Once the LSBG nature is established, galaxies are less prone to significant changes in central surface density and morphology. The inside-out angular momentum redistribution driven by major mergers is part of ongoing research (Pérez-Montaño et al. 2026, in prep.) and aligns with our findings. 

We recall the reader that environmental mechanisms mainly impact satellite galaxies and therefore were excluded from this analysis. By focusing only on central galaxies, we effectively remove those galaxies that are mostly affected by environmental quenching and rapid evolutionary transformations. Therefore, our findings are mainly applicable to central galaxies, where internal mechanisms and halo characteristics play a dominant role, and they may not fully represent the environmental influences experienced by satellite systems \citep{LEPM22, LEPM24}.

Other variations in galaxy properties, such as morphology and star formation rates, are primarily driven by minor mergers, which increase stellar content and star formation in the outskirts while preserving the LSBG nature, In contrast, environmental processes---such as tidal interactions, ram-pressure stripping, and galaxy harassment---and their impact on the evolution of satellite galaxies, as well as systems in groups and clusters, warrant further investigation through high-resolution cosmological simulations and controlled numerical experiments in future studies.

\section*{Acknowledgements}

The authors thank the referee, Hangci Du, for the valuable comments, which have contibuted to improve the quality of this work. The authors acknowledge Gilberto Zavala for facilitating access to computational resources and databases utilized in this work. We also thank Dylan Nelson for facilitating our analysis of IllustrisTNG data via the JupyterLab Workspace. L.E. Pérez-Montaño acknowledges Go Ogiya for his role in providing access to the academic, administrative, and financial resources that supported the development of this research. Luis Enrique Pérez-Montaño was supported by the National Key Research and Development Program of China (No. 2022YFA1602903), the National Natural Science Foundation of China (No. 12373004, W2432003), and the Fundamental Research Fund for Chinese Central Universities (No. NZ2020021, 226-2022-00216). Bernardo Cervantes Sodi acknowledge the financial support provided by PAPIIT project IN111825 from DGAPA-UNAM. D.S. acknowledges the support from the grant CNS2022-135878 funded by MICIU/AEI/10.13039/501100011033 and by the European Union NextGenerationEU/PRTR and thanks the \textit{Flying Bean Coffee Valencia} for their existence. T.-W. C. is supported by the National SKA Program of China (No. 2025SKA0150101). This work is partially supported by cosmology simulation database (CSD) in the National Basic Science Data Center (NBSDC-DB-10).\\ 
In the loving memory of Sandra Montaño Rodriguez\textsuperscript{\textdagger}.

\section*{Data availability}

The data from the IllustrisTNG simulations employed in this work is publicly available at the website \href{https://www.tng-project.org}{https://www.tng-project.org} \citep{Nelson19}.



\bsp	
\label{lastpage}
\end{document}